\documentclass[%
 reprint,
 amsmath,amssymb,
 aps,
]{revtex4-2}

\usepackage{graphicx}
\usepackage{dcolumn}
\usepackage{dsfont}
\usepackage{amssymb}
\usepackage{comment}
\usepackage{xcolor}
\usepackage{bm}
\usepackage{hyperref}
\usepackage{physics}
 \usepackage{booktabs} 

\begin{document}

\preprint{APS/123-QED}

\title{A fault-tolerant encoding for qubit-controlled collective spins}

\author{Charlotte Franke}
\author{Dorian A. Gangloff}%
 \email{dag50@cam.ac.uk}
\affiliation{%
 Cavendish Laboratory, University of Cambridge, JJ Thomson Laboratory, CB3 0US, United Kingdom
}%

\date{\today}

\begin{abstract}
Quantum error correction (QEC) is indispensable for scalable quantum computing, but implementing it with minimal hardware overhead remains a central challenge. Large spin systems with collective degrees of freedom offer a promising route to reducing the control complexity of qubit architectures while retaining a large Hilbert space for fault-tolerant encoding. However, existing proposals for logical gates and QEC in spin ensembles generally rely on inefficient higher-order interactions. Here we introduce spin-$N$-Cat codes, which encode logical qubits in superpositions of spin-coherent states and generalize bosonic Cat codes to the modular subspaces of permutationally symmetric spin ensembles. The code corrects collective and individual dephasing, excitation, and decay errors. We also present an efficient physical realization in central-spin systems, such as a quantum dot, where encoding, decoding, and a universal, fault-tolerant, and bias-preserving gate set are implemented using only first-order interactions. Numerical simulations demonstrate high logical fidelity under dephasing and excitation-decay noise, independent of noise bias, and that full QEC cycles are feasible with realistic microscopic parameters. For the large collective spins available in quantum dots, this translates into a substantial extension of coherence time. Our results establish spin-$N$-Cat codes as a scalable, hardware-efficient approach to QEC in spin-based quantum architectures.
\end{abstract}

                            
\maketitle


\section{Introduction}
Quantum computers promise to solve problems intractable for classical systems, such as factoring large numbers and simulating quantum materials. However, their susceptibility to errors from environmental interactions and imperfect gate operations remains a major challenge. To enable practical quantum computation, robust error correction is essential to maintain coherence and suppress computational errors.  

Classical error correction relies on redundancy to detect and fix errors. In quantum systems, the no-cloning theorem prevents direct duplication of quantum states, requiring a different approach: quantum error correction (QEC) encodes logical qubits into entangled states of multiple physical qubits. This distributes quantum information across the system, allowing errors to be detected and corrected without directly measuring the encoded state. When combined with fault-tolerant gates, which prevent correctable errors from propagating, QEC enables reliable quantum computations as long as the error rate remains below a critical threshold.  

Early QEC codes were developed under the assumption of unstructured noise, where all single-qubit Pauli errors occur with equal probability. These codes were also primarily designed for qubit-based systems, in which quantum information is encoded in two-level subsystems. While theoretically robust, such approaches require significant overhead: multiple physical qubits must be allocated for encoding, syndrome extraction, and correction \citep{Shor.1995, Steane.1995, Calderbank.1996, Steane.1996}. This overhead presents a serious challenge for scaling up quantum hardware with limited qubit resources.

To address this, more recent work has focused on tailoring QEC codes to the structure of physical noise. In systems where dephasing errors dominate over bit-flip errors, for example, biased-noise codes such as the XZZX surface code have demonstrated significantly reduced overhead, with error thresholds improving by an order of magnitude compared to the standard surface code \citep{Ataides.2021}. A complementary strategy is to move beyond two-level systems and make use of multi-level quantum systems (i.e. qudits), where the larger Hilbert space enables more efficient encoding and may simplify gate implementations \citep{Gottesman.2001, Brock.2024}. Both approaches aim to reduce the resource demands of fault-tolerant quantum computing by aligning code structure more closely with the physical characteristics of the system.

In continuous bosonic systems, Cat codes encode quantum information in superpositions of coherent states, exploiting both structured noise and the large Hilbert space of a single oscillator to reduce overhead \citep{Leghtas.2013, Mirrahimi.2014, Lescanne.2020}.  
They are particularly effective in photon-loss–dominated systems, where the logical encoding in even–odd photon-number parity suppresses bit-flip errors exponentially with increasing cat size, while phase errors grow only linearly.
Autonomous stabilization of the codespace via two-photon dissipation and Kerr nonlinearities has yielded long-lived, bias-preserving logical qubits with lifetimes beyond break-even~\cite{Ofek.2016}.  
Recent advances, including concatenated repetition-cat architectures and fast bias-preserving two-qubit gates, further establish cat codes as a hardware-efficient route to fault-tolerant computation~\cite{Guillaud.2019}.

Large spin systems offer a discrete, finite-dimensional analogue of bosonic modes: via the Holstein-Primakoff transformation, the collective operators $(I_x,I_y,I_z)$ map to oscillator quadratures as the total spin length $I\to\infty$ \citep{Holstein.1940}. In this regime, spin-cat codes -- superpositions of oppositely oriented spin-coherent states -- realize the spin-based counterpart of bosonic cat codes, suppressing one dominant error channel through noise-bias alignment. Full fault tolerance, however, requires an additional repetition layer~\cite{Omanakuttan.2024,Kruckenhauser.2025}.

A richer, discrete Hilbert-space structure arises in collective codes, which encode logical information in the symmetric subspace of many spin-$\tfrac{1}{2}$ particles. Representative examples include permutation-invariant multispin Clifford codes, which protect against collective noise using carefully weighted Dicke superpositions, and spin-GKP-type codes, which embed redundancy in grid-like structures in collective-spin phase space. In practice, realizing such encodings typically requires either the preparation of highly structured Dicke superpositions or access to engineered nonlinear collective interactions, such as spin-squeezing dynamics together with high-resolution collective measurements~\cite{Pollatsek.2004,Ouyang.2014,Omanakuttan.20231i,Sharma.2024}. These requirements make their implementation in large ensembles experimentally demanding.

The central-spin system (CSS), where an auxiliary qubit is strongly coupled to a spin ensemble, provides a promising platform for the efficient realization of collective-spin codes. For example, in semiconductor quantum dots (QDs), an electronic spin coherently controls the collective modes of $\mathcal{O}(10^5)$ nuclear spins~\cite{Gangloff.2019}. Recent experiments in QDs have demonstrated all essential ingredients for collective QEC, including electron–nuclear quantum-state transfer~\cite{Appel.2025} with tuneable interaction strength~\cite{Shofer.2024}, single-magnon detection~\cite{Jackson.2021}, ensemble purification via quantum feedback~\cite{Jackson.2022}, and nuclear coherence times up to 100~ms~\cite{Dyte2025}. Moreover, the electronic spin can be used to tune and reduce the spin length $I$ of the collective mode~\cite{Gangloff.2021}, all the way down, in principle, to a many-body singlet state~\cite{Zaporski.2023}. This combination of long nuclear coherence times and fast, conditional electron–nuclear control makes the QD CSS an ideal testbed for collective-spin QEC.

In this paper, we introduce a fault-tolerant family of Cat codes and then show that it is efficiently realisable in the QD-based CSS. The encoding partitions the collective Hilbert space into modular subspaces, enabling simultaneous protection against dephasing and excitation–decay errors within a single code space. The order $N$ sets the modular spacing and can be tuned to match the intrinsic noise bias of the physical system, achieving full fault tolerance without concatenation. Previously proposed spin-2-Cat encodings arise as the lowest-order member of this family. 
In contrast to previous collective codes that depend on carefully engineered Dicke-state superpositions, the spin-$N$-Cat code employs physically meaningful codewords -- symmetric superpositions of spin-coherent states equally spaced along the equator of the generalized Bloch sphere -- that arise naturally under collective control.  
Realized entirely through first-order central-spin couplings already available experimentally, the approach combines modular Hilbert-space partitioning with a hardware-native implementation in a collective-spin system. Using the CSS native physical gates, we construct a universal, bias-preserving logical gate set that operates entirely within the encoded manifold, enabling fault-tolerant operations without breaking the noise bias.
Numerical simulations confirm that the spin-$N$-Cat code satisfies the Knill–Laflamme conditions asymptotically and maintains high logical fidelity under both dephasing and excitation–decay noise. To quantify intrinsic robustness, we extract the maximum interval over which the system can evolve uncorrected while still allowing successful recovery.  
Compared to a naive Dicke-state superposition encoding~\cite{Appel.2025}, the spin-$N$-Cat code extends the coherence time by up to $15\times$.  
This level of passive protection is comparable to that predicted for bosonic cat codes ~\cite{Leghtas.2013}, demonstrating that modular partitioning of collective-spin Hilbert spaces can achieve both practical performance and experimental feasibility.  

Beyond a single ensemble, the same principles can be extended to encoding information in multiple nuclear-spin modes~\cite{Appel.2025}, or to coupling multiple central-spin nodes to enable networks of protected collective qudits for scalable quantum architectures.

\section{Spin \textit{N}-Cat Code}
\begin{figure*}
    \centering
    \includegraphics[width=\linewidth]{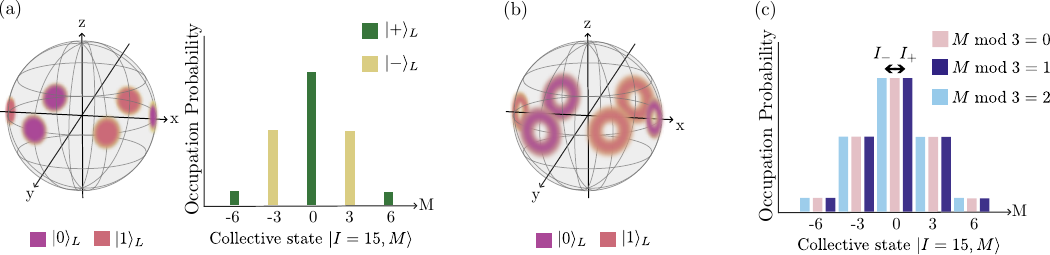}
    \caption{
    (a) Left: Wigner distribution of the code states on the generalized Bloch sphere. These states are superpositions of three coherent states on the equator, separated by an angle of \( \frac{2\pi}{3} \). Right: Their probability distribution in the collective basis lies in the \( M \bmod{3} = 0 \) subspace. 
    (b) Effect of a dephasing-type error (\(I_z\)). The Wigner distribution broadens, but the logical states remain non-overlapping. (c) Effect of an excitation or decay error (\(I_\pm\)), which shifts the \( M \bmod{3} \) value -- an error syndrome.
}
    \label{Figure_1}
\end{figure*}
We now introduce the spin-\(N\)-Cat encoding and show how its structure ensures fault tolerance against physically relevant errors.
The system consists of an ensemble of identical spin-\(\tfrac{1}{2}\) particles described by collective angular momentum operators
\(
I_\alpha = \tfrac{1}{2}\sum_n \sigma_{n,\alpha}
\).
Under collective operations, the dynamics remain fully permutationally symmetric and can be expressed in the collective-spin basis
\( \ket{I,M} \), where \( I \) denotes the total angular momentum and
\( M \in [-I,I] \) its projection along the quantization axis \( z \)
\citep{Dicke.1954, Chase.2008, Shammah.2018}, so that $I_z = \sum_{M=-I}^{I} M \ket{I,M}\bra{I,M}$.
Upon restriction to a fixed total-spin sector \(I\), the collective operators
\( I_\alpha \) act irreducibly and reduce to the standard spin-\(I\) angular-momentum generators, obeying the usual ladder relations.
Within each \(I\) sector, these states may be represented on a generalized Bloch sphere, whose poles correspond to the fully polarized states \( M=\pm I \).

The dominant errors in the system are collective dephasing
(\(I_z\)) and collective excitation and decay
(\(I_+\) and \(I_-\)) processes. We define the set of correctable errors as
\begin{equation}
\begin{aligned}
    \mathcal{E}_{k,l} := &\text{span} \Big\{ I_{\alpha_1} I_{\alpha_2} \cdots I_{\alpha_n} \;\Big|\; I_{\alpha_j} \in \{ I_+, I_-, I_z \},\\& \#I_+ \leq k,\ \#I_- \leq k,\ \#I_z \leq l \Big\},
\end{aligned}
\end{equation}
where each element of \(\mathcal{E}_{k,l}\) represents an error process
involving at most \(k\) collective spin-flip events and \(l\) collective
phase-error events.
The order \(N\) of the Cat code determines the number of correctable excitation-type errors and is chosen to match the noise structure of the physical system under consideration.

To construct this encoding, we draw inspiration from bosonic Cat codes, which encode logical states as superpositions of coherent states spaced symmetrically in phase space. In the 4-Cat code, for example, the codespace is supported on four coherent components and restricted to Fock states with photon numbers $n \equiv 0 \pmod{4}$, enabling photon loss to be detected as a change in parity~\citep{Mirrahimi.2014, Ofek.2016}. At the same time, their phase-space separation protects against dephasing by maintaining distinguishability \citep{Puri.2020, Lescanne.2020, Guillaud.2019}.

We adopt a similar strategy in the spin ensemble: logical states are constructed as superpositions of spin-coherent states placed symmetrically along the equator, combining phase-space separation with support confined to a modular subspace \( M \bmod (N/2) = 0 \). Formally, the spin-\(N\)-Cat logical states are defined as
\begin{equation}
\begin{aligned}
    \ket{0}_L &\propto \sum_{i=1}^{N/2} \ket{I, \tfrac{\pi}{2}, \tfrac{4\pi i}{N}}, \\
    \ket{1}_L &\propto \sum_{i=1}^{N/2} \ket{I, \tfrac{\pi}{2}, \tfrac{4\pi i}{N} + \tfrac{2\pi}{N}},
\end{aligned}
\end{equation}
where \( \ket{I, \theta, \phi} \) denotes a spin-coherent state with polar angle \( \theta \) and azimuthal angle \( \phi \) on the Bloch sphere.  The Hilbert space thus decomposes into \( N/2 \) orthogonal sectors distinguished by the modular value of \( M \), each serving as an error-syndrome space analogous to the Fock-parity subspaces in bosonic Cat codes.  The case \( N = 2 \) reduces to the spin-2-Cat code~\citep{Omanakuttan.2024}.  
Excitation or decay processes \( I_\pm \) change the modular label and thereby move the state between sectors, providing an error syndrome and enabling correction of up to \( k = (N - 2)/4 \) such errors.

To visualize this structure, consider the 6-Cat code shown in Fig.~\ref{Figure_1}. The Wigner distribution shows the logical states each as superpositions of $3$ spin-coherent states placed symmetrically along the equator (Fig.~\ref{Figure_1}a, left panel). In the collective spin basis, the logical states form a binomial distribution centered at \( \ket{I, M = 0} \), with support restricted to the \( M \bmod{3} = 0 \) subspace (Fig.~\ref{Figure_1}a, right panel). Dephasing errors broaden the Wigner distribution but leave the states distinguishable (Fig.~\ref{Figure_1}b), demonstrating the code’s resilience to phase noise. Collective excitations or decays shift the state into orthogonal subspaces with \( M \bmod{3} = 1 \) or \( 2 \), enabling error correction based on syndrome detection (Fig.~\ref{Figure_1}c).

Dephasing errors, generated by powers of \(I_z\), act within a single modular sector.  
Since each sector has dimension \(\sim 4I/N\), the two logical states remain distinguishable under polynomial functions of \(I_z\) up to order \(\mathcal{O}(2I/N)\).  
Numerically, we find
\[
\langle 0_L | I_z^m I_z^n | 1_L \rangle \approx 0 
\quad \text{for all } m,n \le l = \alpha(N) I + \beta(N),
\]
with \(\alpha \approx 2/N\), consistent with this dimensional bound.  
The small offset \(\beta(N)\) reflects that the encoding does not occupy the full Hilbert space, as it is optimized for physical realizability with first-order collective interactions rather than for saturating the theoretical limit.  Consequently, the Knill–Laflamme conditions~\citep{Knill.1997} are satisfied asymptotically: excitation and decay processes transfer population between orthogonal modular sectors, while pure dephasing acts within one sector.  
Mixed errors, which combine dephasing with raising or lowering, can also change the modular label but perturb the intra-sector phase structure only weakly, so the logical overlap remains suppressed to leading order.  
A detailed derivation, including the dimension-scaling argument, geometric interpretation, and numerical validation, is provided in Appendix~\ref{sec:KL_appendix}.

Finally, while the discussion so far has assumed collective coupling to the environment, real systems also experience local decoherence, where each spin interacts with its own environment.  
Single-spin raising or lowering events then change the total angular momentum \(I\), causing leakage out of the original codespace.  
In the large-\(I\) limit, however, the logical encoding depends primarily on the symmetry of the coherent components rather than the exact value of \(I\), so once \(I\) is identified the same error set \(\mathcal{E}_{k,l}\) applies, allowing correction of both collective and individual noise processes.

\section{Implementation in a Central Spin System} \label{CSS_Implementation}
\begin{figure*}
    \includegraphics[width=\linewidth]{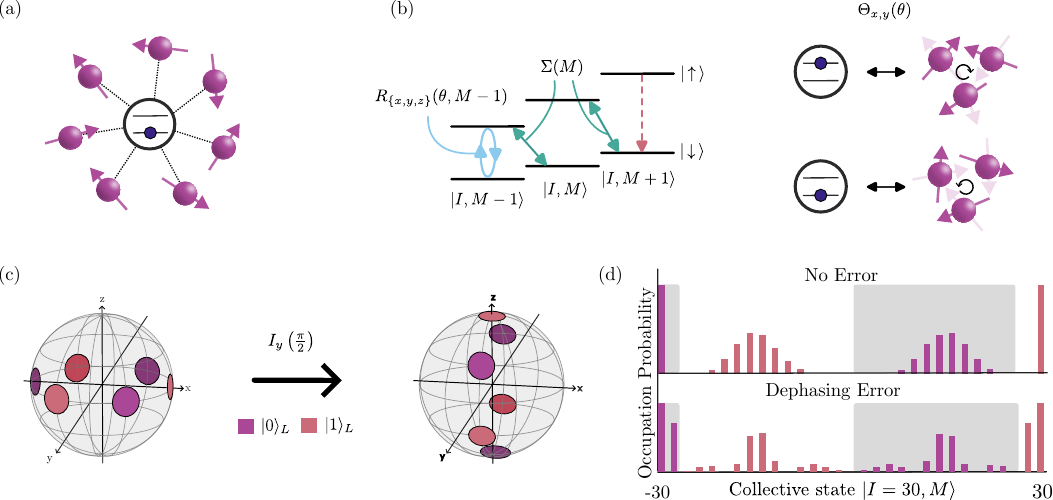}
    \caption{
    (a) Physical system: the ensemble interacts with a single auxiliary spin, enabling nonlinearity and readout. 
    (b) Available gates: unconditional qubit rotations \( R_i(\theta) \), conditional rotations \( R_{\{x,y,z\}}(\theta, M) \), ensemble rotations \( \Theta(\theta) \), and flip-flop transitions \( \Sigma(M, t) \). 
    (c) The logical states \(\ket{0}_L\) and \(\ket{1}_L\) consist of spin-coherent components at different azimuthal angles. After a \(\pi/2\) rotation about the \( y \)-axis, these states are mapped to different latitudes on the Bloch sphere, leading to distinct projections in \( I_z \).  
    (d) Probability distributions in the \( I_z \)-basis before and after a dephasing error. The rotated logical states have minimal overlap and remain distinguishable even under noise, enabling fault-tolerant conditional control.
    }
    \label{fig:physical_implementation}
\end{figure*}

In this section, we describe how the $N$-Cat code can be implemented in a central-spin architecture. We introduce the physical gate primitives, show how they generate a universal, fault-tolerant, bias-preserving logical set, and discuss how the same operations implement error correction.

\subsection{Hamiltonian and native gates}

A central-spin system consists of a single spin coupled to a large ensemble of surrounding spins (Fig.~\ref{fig:physical_implementation}a) \cite{Urbaszek.2013}. When the couplings are approximately permutation-invariant, the ensemble dynamics are well described in terms of collective operators $\vec I$, forming symmetric states with large effective spin length. The central spin acts as a controllable mediator, enabling indirect manipulation of these collective degrees of freedom without individual spin addressing \cite{Jackson.2021}.

We work with a driven central-spin architecture in which the effective interaction can be taken to have the form
\begin{equation}
H =
\Omega S_x 
+ \delta S_z
+ \omega_n I_z
+ a\, S_z I_z
+ a_{nc}\, S_z I_x ,
\label{eq:H_eff}
\end{equation}
where $\Omega$ defines the central-spin drive, $\delta$ is the detuning, and $a$ and $a_{nc}$ denote longitudinal and transverse interactions. The term $a S_z I_z$ produces ensemble-dependent frequency shifts, while $a_{nc} S_z I_x$ enables conditional transverse rotations and drive-assisted sideband transitions. This effective Hamiltonian provides the control structure underlying the gate primitives introduced below; Appendix~C gives an explicit construction.

All operations required for encoding, decoding, and fault-tolerant operation of the $N$-Cat code can be built from five elementary unitary transformations acting on the central spin (the qubit) and the ensemble:

\begin{enumerate}
    \item The \emph{conditional ensemble rotation}, 
    \[
    \Theta(\theta) = \exp(-i \theta S_z I_x),
    \]
    rotates the collective spin about the $x$-axis, conditioned on the central-spin state.

    \item The \emph{conditional qubit rotation},
    \begin{equation*}
        R_{\{x,y,z\}}(\theta, M) = \exp\left(-i \frac{\theta}{2} S_{\{x,y,z\}} \ket{I,M}\bra{I,M}\right),
    \end{equation*}
    applies a single-qubit rotation to the central spin along the $x$, $y$, or $z$ axis, conditioned on the ensemble being in the collective state $\ket{I, M}$.

    \item The \emph{conditional flip-flop and flip-flip transitions} exchange excitations between the central spin and the ensemble within a selected $M$-subspace,
    \[
    \Sigma_{\text{flip-flop}}(t, M) = \exp\left(-it [S_+ I_- + S_- I_+] \ket{I, M}\bra{I, M}\right),
    \]
    \[
    \Sigma_{\text{flip-flip}}(t, M) = \exp\left(-it [S_+ I_+ + S_- I_-] \ket{I, M}\bra{I, M}\right),
    \]
    thereby modifying the collective angular momentum by one unit.

    \item The \emph{unconditional single-qubit rotations}
    \[
    \mathcal{X}(\theta) = \exp\left(-i \frac{\theta}{2} S_x\right), 
    \qquad 
    \mathcal{Y}(\theta) = \exp\left(-i \frac{\theta}{2} S_y\right),
    \]
    act independently of the ensemble state. 

    \item The \emph{unconditional ensemble rotations} are global collective rotations
    \[
    I_{x,y}(\theta),
    \]
    applied uniformly across the spin ensemble.
\end{enumerate}

Gates 1--4 can be realized by tuning Hamiltonian Eq.~\ref{eq:H_eff} through appropriate drive parameters and control sequences. Conditional qubit rotations are obtained by choosing the drive frequency such that the effective detuning $\delta + a I_z$ vanishes for a selected collective state, bringing the central spin into resonance only within that $M$-subspace. Conditional flip-flop and flip-flip transitions are activated in the sideband regime ($\Omega \ll \delta$). In this limit, the transverse interaction $a_{nc} S_z I_x$ mediates excitation exchange between the central spin and the ensemble. Resonance occurs when the detuning matches the energy cost of changing the collective spin projection, enabling selective transitions $\ket{I,M} \leftrightarrow \ket{I,M\pm1}$ conditioned on the central-spin state (Fig.~\ref{fig:physical_implementation}b). The same transverse interaction can be harnessed to implement conditional ensemble rotations. By applying periodic pulse sequences that modulate the sign of $S_z$, the dominant longitudinal contribution can be averaged out, isolating an engineered $S_z I_{x/y}$ interaction.

Beyond the physical gate primitives described above, we require a mechanism to distinguish logical states through their collective observables, as this is essential for syndrome detection and conditional control. To this end, we apply a $\pi/2$ rotation about the $y$-axis (Fig.~\ref{fig:physical_implementation}c), $I_y(\pi/2)$. In the original basis, both logical states share the same $I_z$ distribution and differ only by relative phase. After rotation, their orthogonality manifests as distinct $I_z$ support (Fig.~\ref{fig:physical_implementation}d), corresponding geometrically to a shift from azimuthal separation on the equator to separation in latitude. Numerically, we find that the rotated logical and error states exhibit well-separated $I_z$ support with exponentially suppressed overlap as $I$ increases (App.~\ref{sec:KL_appendix}). This separation enables fault-tolerant control: conditional electron operations can target a given logical state or error space without mixing them, even in the presence of dephasing noise.

In the rotated frame, collective dephasing induces phase shifts between different $I_z$ sectors, which can reduce the effective noise bias during control. We therefore operate on timescales short compared to the dephasing time, ensuring that accumulated phases remain small and bias-preserving control is maintained.
\subsection{Encoding and Decoding}
\begin{figure}
    \centering
    \includegraphics[width=\linewidth]{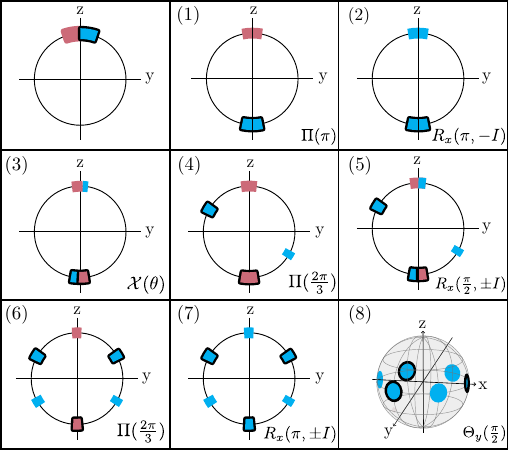}
  \caption{
    Sequence of operations for encoding and decoding the state of the electron into the codespace of the 6-Cat code. The joint electron–ensemble state is represented by a generalized Fresnel diagram: position indicates the collective state of the ensemble, color indicates the electron state (blue for ground, red for excited). Black circles mark states that begin with the electron in the ground state and, as required, end in the logical state \(\ket{0}_L\). In step~3, we use \(\theta = 2\arccos\!\left(\sqrt{\tfrac{2}{3}}\right)\).}
    \label{fig:Encoding}
\end{figure}

We build on the algorithm of Leghtas et al.~for encoding a 2-Cat state in a bosonic mode \citep{Leghtas.2013}. In our implementation, the encoding is created in the rotated codespace, where logical states are distinguishable in the \(I_z\)-basis and conditional operations can be applied, and in the final step the state is rotated such that the coherent components lie on the equator. The 6-Cat sequence is shown in Fig.~\ref{fig:Encoding}. We define \(\Pi(\phi)\) as a combination of \(\Theta\) and an unconditional ensemble rotation, effectively rotating the ensemble only if the electron is in the ground state. Starting from \(\ket{I,I}\otimes(\alpha\ket{\downarrow}+\beta\ket{\uparrow})\), the \(\Pi(\pi)\) gate (step~1) and conditional qubit rotation \(R_x(\pi,-I)\) (step~2) bring the system into \(\ket{\downarrow}(\alpha\ket{I,I} + \beta\ket{I,-I})\), and the \(\mathcal{X}(\theta)\) gate disentangles the electron (step~3), producing a 2-Cat code along the \(z\)-axis.  For the general spin-\(N\)-Cat code, this 2-Cat state is the initial state for the subsequent steps. Alternating conditional ensemble and qubit rotations then generate a superposition of \(N\) spin-coherent states in the \(I_x=0\) plane:
\[
\prod_{i=1}^{N/2} R_x\!\left(2\sqrt{\frac{N/2 - i}{N/2 - i + 1}}, \pm I\right)\, \Pi\!\left(\frac{2\pi}{N}\right).
\]
A final ensemble rotation \(\Theta_y(\pi/2)\) (step~8 for the 6-Cat code, Fig.\,\ref{fig:Encoding}) completes the encoding; decoding is achieved by reversing the sequence. 

\subsection{Logical Gates}
\begin{figure*}
    \centering
    \includegraphics[width=\linewidth]{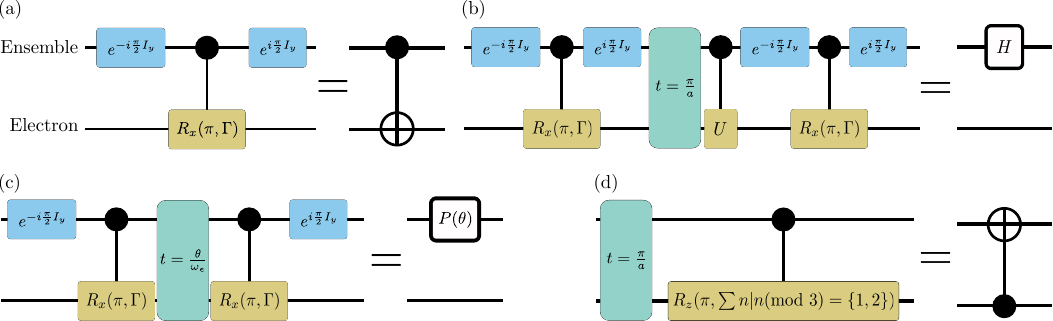}
    \caption{
    Fault-tolerant logical gates for the 6-Cat code. 
    In all circuits, the top line represents the ensemble (logical qubit) and the bottom line the central spin (electron).  
    (a) CNOT with ensemble as control.  
    (b) CNOT with electron as control.  
    (c) Logical Hadamard, implemented from a CNOT (ensemble control), free evolution under \(A_c S_z I_z\), a corrective single-qubit unitary \(U\), and a second CNOT.  
    (d) Logical phase gate \(P(\theta)\), realised by entangling the electron and ensemble, accumulating a controlled phase during free evolution, and mapping the phase back to the ensemble.  
    All gates act identically on the codespace and all correctable error spaces, preserving \(M \bmod \tfrac{N}{2}\) and thus the noise bias.
    }
    \label{fig:gates_sq}
\end{figure*}
We implement a universal logical set
\[
\{\mathrm{CNOT},\,H,\,T\},\qquad T=P(\pi/8),
\]
using the physical controls introduced above. All three gates are implemented in the rotated frame, where the logical states are separated in the $I_z$ basis and can be addressed through collective control. In this frame, however, the dynamics are sensitive to $I_z$ dephasing, so maintaining the bias during control requires careful pulse design.

Bias preservation ensures that collective dephasing errors are not converted into raising or lowering errors. We distinguish two levels of bias preservation. 
At the gate level, a logical operation is bias-preserving if its overall action commutes with the dominant noise channel, so that dephasing-type errors remain dephasing on the logical subspace. 
For the spin-$N$-Cat code, this corresponds to preserving the modular polarization $M \bmod N/2$ during the logical map. 
At the Hamiltonian level, the instantaneous control Hamiltonian must commute with $I_z$ at all times, maintaining bias preservation continuously throughout the physical evolution. 
The gates below satisfy bias preservation at the gate level but not strictly at the Hamiltonian level: intermediate steps in the rotated frame may transiently couple different $I_z$ sectors. Such excursions do not significantly affect the logical bias provided the gate duration is short compared with the dephasing time.

Each gate must therefore (i) implement the correct logical action on the codespace, (ii) act consistently across all correctable error spaces~\citep{Gottesman.2000}, and (iii) preserve the biased noise structure by maintaining $M \bmod \tfrac{N}{2}$.

Let \(\Gamma\) denote the set of collective projections that support \(\ket{0}_L\) together with its correctable error spaces in the rotated frame (for the 6 Cat code this set is the grey region in Fig.~2d), and $R_x(\pi,\Gamma)$ the electronic inversion thus conditioned on \(\Gamma\). Figure \ref{fig:gates_sq} shows the logical set construction in terms of physical gates, as follows, and further details are provided in Appendix~\ref{Appendix_gates}.

\paragraph{CNOT with ensemble control (C$_n$NOT$_e$).}
A conditional electron \(\pi\) rotation in the rotated code space restricted to \(\Gamma\) flips the electron exactly when the ensemble occupies the logical \(\ket{0}_L\) sector. An ensemble rotation \(I_y(-\pi/2)\) then returns to the original codespace. Because the \(I_z\) supports of \(\ket{0}_L\) and \(\ket{1}_L\) are disjoint in the rotated frame, the gate acts identically on the codespace and on the corresponding error spaces, and the ensemble polarization, hence \(M \bmod \tfrac{N}{2}\), is unchanged.

\paragraph{Hadamard.}
We first apply a CNOT with ensemble control to entangle the electron and the ensemble. The system then evolves under \(a S_z I_z\) for \(t=\pi/a\), after which a single qubit electron unitary \(U\) that depends only on \(M \bmod \tfrac{N}{2}\) cancels the relative phase. A second CNOT with ensemble control disentangles the qubits. Each step either targets \(\Gamma\) or depends only on \(M \bmod \tfrac{N}{2}\), which ensures the same logical action on all correctable error spaces and preserves \(M \bmod \tfrac{N}{2}\).

\paragraph{Phase gate.}
A logical phase gate \(P(\theta)\) is obtained by mapping a controllable electron Zeeman phase onto the ensemble: a conditional electron \(\pi\) rotation acting only on \(\Gamma\) first entangles the electron and the ensemble. 
During subsequent free evolution under \(\omega_e S_z\) for \(t=\theta/\omega_e\), the electron acquires a dynamical phase \(e^{-i\theta S_z}\). 
A second conditional electron \(\pi\) rotation then disentangles the two systems, transferring this accumulated phase to the ensemble through phase kickback. 
An ensemble rotation \(I_y(\pi/2)\) restores the original logical basis. 
Setting \(\theta=\pi/8\) yields the \(T\) gate. As with the other gates, \(\Gamma\) selective control together with corrections that depend only on \(M \bmod \tfrac{N}{2}\) ensures fault tolerance and preserves the noise bias.

While not required for universality, the reverse-direction CNOT can also be implemented natively in our platform, and does not involve a transition to the rotated frame.
\paragraph{CNOT with reversed control (C$_e$NOT$_n$).}
Using free evolution under the collinear hyperfine interaction for \(t=\pi/a\) flips the logical ensemble state if the electron is \(\ket{\uparrow}\). If an $I_\pm$ error shifts the collective magnetization, the electron acquires an extra phase of \(\pi\), which we remove with an electron rotation \(R_z(\pi)\) conditioned only on \(M \bmod \tfrac{N}{2}\). With this modular correction the operation is uniform across all relevant subspaces and preserves \(M \bmod \tfrac{N}{2}\).

\subsection{Error Correction} 
\label{subsec:ErrorCorrection}
\begin{figure*}
    \includegraphics[width=\linewidth]{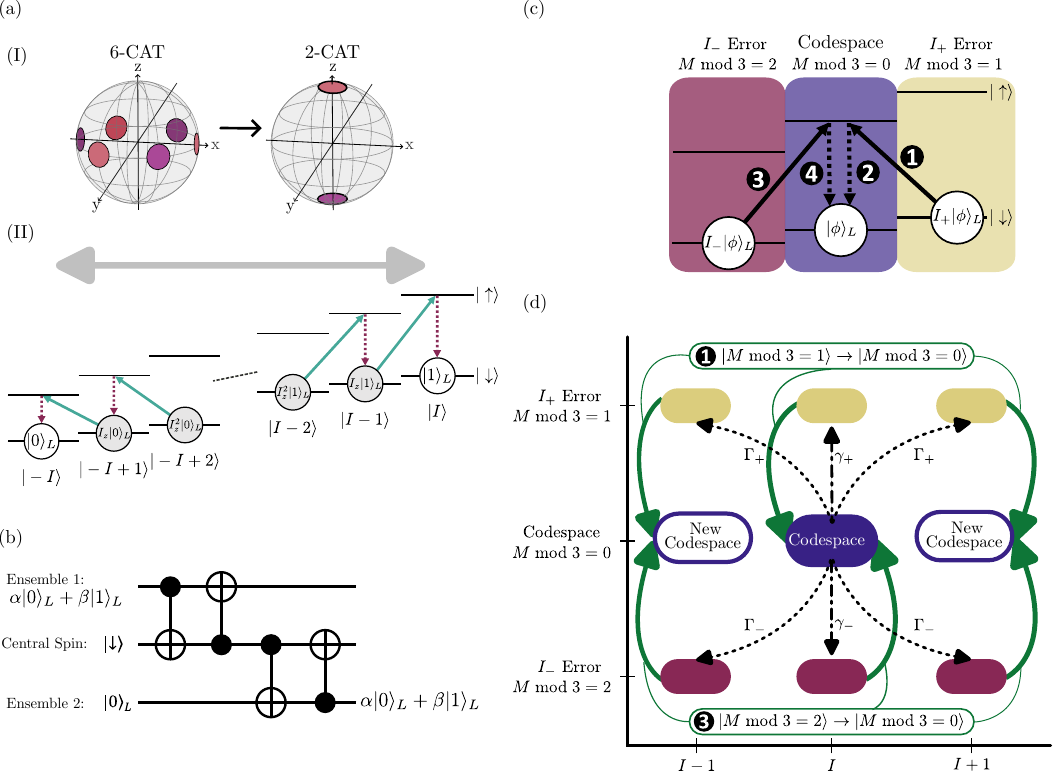}
    \caption{Error Correction for the 6-Cat code. (a) Correction of collective dephasing errors. (I) The 6-Cat encoding is first mapped onto a 2-Cat encoding along the z-axis. Dephasing errors are now observable as reduced polarization of the coherent states. (II) A directional pumping process, implemented through alternating flip-flop and electron reset gates, restores the polarization and corrects dephasing errors. (b) Alternative method for correcting dephasing errors: The fault-tolerant CNOT gate transfers the information to the electron qubit, which is then used to encode the information onto a new ensemble. (c) Correction of decay or pumping errors: Selective flip-flop transitions bring the \(M \bmod 3 = 1, 2\) subspaces back into the codespace, restoring the encoded information. (d) Collective vs. Local Errors: A local error alters the total angular momentum \(\mathbf{I}\), but since the gate \(\Sigma\) operates independently of \(\mathbf{I}\), it corrects both collective and local errors. While a local error changes the total angular momentum of the codespace, all other properties remain unaffected.}
    \label{fig:error_correction}
\end{figure*}
To illustrate the principles of error correction, we focus on the 6-Cat code; higher-order Cat codes operate analogously. We distinguish between the correction of raising/lowering errors ($I_\pm$) and dephasing errors ($I_z$). For $I_\pm$-errors, we present a correction method that operates within one collective spin system and can be implemented autonomously using reinitialization of the central spin. For $I_z$-errors, we present two alternative schemes: one that operates autonomously within a single ensemble, and one that transfers the logical state to another ensemble via the fault-tolerant CNOT gate. This latter approach is naturally suited to quantum dot systems with two distinct nuclear species, which can be independently addressed due to differences in their Zeeman energies and hyperfine couplings\cite{Shofer.2024,Appel.2025}.

As discussed earlier, correcting raising and lowering errors requires returning the system from the \(M \bmod 3 = 1,2\) subspaces back to the codespace \(M \bmod 3 = 0\). This is achieved by applying \(\Sigma_{\text{flip-flop}}(t_M, M)\) 
with interaction times \(t_M = \pi/(2 g_M)\), 
where \(g_M\) is the collective matrix element, 
thereby mapping
\[
\ket{M \bmod 3 = 1} \otimes \ket{\downarrow}
\rightarrow
\ket{M \bmod 3 = 0} \otimes \ket{\uparrow}.
\]
Reinitializing the central spin and then correcting lowering errors using the same approach completes the autonomous correction, as shown in Fig.~\ref{fig:error_correction}c.

We now turn to $I_z$-error correction. The first method is an autonomous single-ensemble scheme in which dephasing errors are corrected by temporarily mapping the 6-Cat encoding to a 2-Cat code along the \(z\)-axis. This is done by reversing the encoding sequence shown in the second and third rows of Fig.~\ref{fig:Encoding}, resulting in the logical basis:
\begin{equation}
    \ket{0^\prime}_L = \ket{I, I}, \qquad \ket{1^\prime}_L = \ket{I, -I}.
\end{equation}
To avoid mapping correctable dephasing errors to logical errors during this process, all conditional qubit rotations are applied resonantly to the \(l\) most polarized states, with $l$ the number of correctable dephasing errors: \(R_x(\pi, \pm I) \rightarrow \sum_{i=0}^l R_x(\pi, \pm I \mp i)\).

In this encoding, dephasing errors now appear as reduced spin polarization. Similar to the correction method used in the spin 2-Cat code \cite{gross2024hardware}, a directional pumping process restores polarization by driving population towards \(\ket{\pm I}_z\). This is implemented using selective flip-flop gates \(\Sigma(t, M)\), combined with electron reinitialization to maintain the directionality of the process (Fig.~\ref{fig:error_correction}a(II)). Finally, the encoding sequence is re-applied to return to the 6-Cat code. This method operates autonomously but has the disadvantage that during the intermediate $2$-Cat step the information is stored in the \(I_z\)-basis and is thus more vulnerable to further dephasing.

As an alternative $I_z$-correction strategy, compatible with systems such as quantum dots which contain multiple spin ensembles~\cite{Appel.2025}, the logical state can be transferred to a new ensemble using fault-tolerant CNOT gates, following an approach similar to the 2-Cat protocol introduced by Omanakuttan et al.~\cite{Omanakuttan.2024}. The procedure begins with a CNOT gate using the original (error-corrupted) ensemble as the control, transferring the logical state onto the central spin. A second CNOT disentangles the ensemble and electron, leaving the now-corrected information encoded in the central spin alone. Subsequently, two additional CNOT operations copy the logical state to a freshly initialized ensemble in \(\ket{0}_L\), completing the state transfer (Fig.~\ref{fig:error_correction}b).

The main advantage of this two-qudit scheme is that it removes the scaling limitation present in single-qudit $I_z$-correction. In the single-qudit protocol, correcting $I_z$-type errors requires a sequence of conditional operations whose number grows with $I$, leading to longer execution times in large ensembles. By contrast, the two-qudit method replaces this stage with a fixed-length state transfer: two CNOTs to map the state from the first ensemble to the electron, and two CNOTs to write it into the second ensemble. Because the number of CNOT operations is fixed and does not depend on $I$, the overall $I_z$-correction time is independent of ensemble size. This scaling behaviour makes the method particularly attractive for large-$I$ systems, where collective enhancement shortens the duration of each CNOT and thus the total correction time.

Finally, as noted earlier, single-spin (as opposed to collective) errors can change the total angular momentum \(I\), causing the system to leak out of the original codespace. The total spin is not easily measured non-destructively, but the measurement of \(M \bmod \frac{N}{2}\), which determines the error syndrome, can be performed independently of \(I\). This allows error correction to proceed without precise knowledge of the total spin. However, an uncertainty in \(I\) effectively reduces the number of correctable errors. For instance, implementing logical gates such as the CNOT requires extending the support set \(\Gamma\) to include relevant states from multiple \(I\) sectors. As shown in Fig.~\ref{fig:error_correction}d, individual spin errors cause leakage out of the codespace, but the same correction procedures used for collective errors successfully return the system to a new, nearly equivalent codespace with \(M \bmod 3 = 0\).

\section{Numerical Simulation}

In this section, we numerically investigate the performance of the spin-$N$-Cat code. We first simulate ideal error-correction cycles with perfect gates to benchmark different $N$-Cat codes across a range of noise-bias configurations, and compare to a two-level Dicke baseline. We then assess the fault tolerance of the logical operations under gate-level errors and experimentally motivated imperfections.

\subsection{Idealized Simulation in a Collective Spin System}

\begin{figure}
    \centering
    \includegraphics[width=\linewidth]{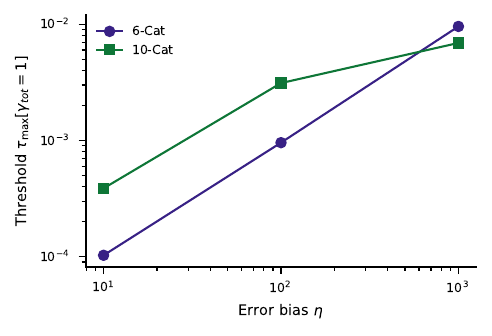}
    \caption{Threshold cycle times $\tau_{\mathrm{max}}$ for different Cat codes at varying noise bias $\eta$ (I=210).}
    \label{fig:sim_ideal}
\end{figure}

To probe the intrinsic performance of the $N$-Cat code against decoherence, we simulate its evolution using the Lindblad master equation,
\begin{equation}
\dot{\rho} = -\frac{i}{\hbar}[H, \rho] + \sum_{m = -1}^{1} \left( \frac{\gamma_m}{2} \mathcal{L}_{I_m}[\rho] + \frac{\Gamma_m}{2} \sum_{n=1}^N \mathcal{L}_{\sigma_m}[\rho] \right),
\end{equation}
where $m \in \{-1, 0, 1\}$ denotes decay ($I_-$/$\sigma_-$), dephasing ($I_z$/$\sigma_z$), and excitation ($I_+$/$\sigma_+$) processes. The rates $\gamma_m$ and $\Gamma_m$ quantify collective and individual decoherence, respectively. In the idealized benchmark below we set $\Gamma_m=0$ and retain only the collective terms, which dominate in the large-spin regime and substantially reduce numerical complexity. As discussed above, correcting collective noise captures the essential behavior of local errors as well.

To quantify performance, we compute the logical fidelity following an error-correction cycle. The protocol proceeds by initializing a logical ensemble qubit, evolving under noisy dynamics for time $\tau$ and then applying the correction circuit. During the idle interval $\tau$ we set $H=0$ and evolve under the Lindblad terms only. We then evaluate the overlap with the original input state. To account for arbitrary logical inputs, we repeat this procedure for both the logical $\ket{0}_L$ state and the logical $\ket{+}_L$ state, and use these to approximate the Bloch-sphere average fidelity. In other words, this simulation evaluates the single-cycle tolerance time of the Cat encoding: the maximum idle interval $\tau$ over which the logical qubit can evolve unprotected 
before a single ideal error-correction cycle is applied. By scanning over $\tau$, we extract the maximum cycle duration $\tau_{\mathrm{max}}$ for which the logical fidelity is maintained above a threshold of $99.9\%$, corresponding to an effective error rate below $10^{-3}$. We do not model repeated correction cycles or gate imperfections, so $\tau_{\mathrm{max}}$ should be interpreted as a per-cycle tolerance bound, not a logical lifetime under continuous error correction.

Figure~\ref{fig:sim_ideal} shows $\tau_{\mathrm{max}}$ for various noise bias configurations, characterized by the parameter $\eta = \gamma_z / (\gamma_+ + \gamma_-)$, with the normalization $\gamma_\text{tot}= \gamma_z + \gamma_+ + \gamma_- = 1$. For each value of $\eta$, the dephasing-recovery order $l$ is chosen to maximize the post-correction logical fidelity. Since we normalize the total noise rate $\gamma_\text{tot}$ to unity, all cycle times are expressed in units of the inverse decoherence rate. As expected, higher-order Cat codes exhibit improved performance in regimes where excitation and decay dominate over dephasing. This improvement follows from the fact that an $N$-Cat code can correct up to $k=\tfrac{N-2}{4}$ raising/lowering errors, so increasing $N$ directly increases robustness to these processes. This highlights the advantage of increasing $N$ in suppressing logical errors when dephasing is not the primary noise mechanism.

\subsection{Comparison to Dicke encoding}

To contextualize this improvement, we compare to a two-level Dicke encoding without error correction, defined by the states $|0\rangle = |I, -I\rangle$ and $|1\rangle = |I, -I{+}1\rangle$. In this scheme, dephasing leads only to slow phase damping (at rate $\gamma_z/2$), while raising or lowering events $I_{\pm}$ cause immediate logical flips or leakage, with rates collectively enhanced by a factor $2I$. The resulting average infidelity scales as $r_{\mathrm{avg}}(t) \simeq \gamma_z t / 6 + 2I(\gamma_+ + \gamma_-) t$, and the corresponding Dicke coherence time is approximately
\[
t_{\mathrm{Dicke}}(\eta)
\simeq \frac{\varepsilon(1+\eta)}{\eta/6 + 2I},
\]
as derived in Appendix~\ref{app:t_dicke}.

Using the numerically extracted $\tau_{\mathrm{max}}$, we define the  improvement
\[
\mathcal{R}(N,\eta) = \frac{\tau_{\mathrm{max}}(N,\eta)}{t_{\mathrm{Dicke}}(\eta)}.
\]
Table~\ref{tab:improvement_factors} reports $\mathcal{R}$ for various values of $N$ and $\eta$. Across the bias regimes considered, the Cat codes achieve a $5$--$15\times$ increase in the single-cycle tolerance time $\tau_{\mathrm{max}}$ relative to the Dicke baseline. This magnitude is qualitatively consistent with the improvement observed for bosonic cat encodings relative to bare qubits in related noise settings \cite{Leghtas.2013}. Here $\mathcal{R}$ should be interpreted as a single-cycle tolerance metric, and can be extended to longer logical lifetimes when embedded into repeated QEC protocols.

\begin{table}[t]
  \centering
  \caption{
Coherence improvement $\mathcal{R}(N,\eta)$ of the spin-$N$ Cat code
relative to a two-level Dicke encoding
($|I,-I\rangle, |I,-I{+}1\rangle$) for a target logical fidelity
$F\!\ge\!0.999$ ($\varepsilon=10^{-3}$).
The improvement is defined as
$\mathcal{R}=\tau_{\mathrm{max}}/t_{\mathrm{Dicke}}$
(see Appendix~\ref{app:t_dicke}).
}
  \vspace{1mm}
  \renewcommand{\arraystretch}{1.15}
  \setlength{\tabcolsep}{6pt}
  \begin{tabular}{lccc}
    \toprule
    & $\eta=10$ & $\eta=100$ & $\eta=1000$ \\
    \midrule
    $N=6$  & 3.93 & 4.13 & 5.60 \\
    $N=10$ & 14.82 & 13.45 & 4.01 \\
    \bottomrule
  \end{tabular}
  \label{tab:improvement_factors}
\end{table}

In addition to the idealized single-cycle analysis above, we test the fault tolerance of the logical gate constructions by inserting Kraus errors prior to each gate, followed by a correction cycle and decoding back onto the electron qubit. We then evaluate the fidelity of the recovered central spin state relative to the original logical input. The 6-Cat code maintains high logical fidelity in the presence of multiple dephasing errors and single excitation or decay events, confirming the fault tolerance of the gate set. For larger ensembles ($I=60$), all gates remain correct up to the expected order of correctable errors, with fidelity above $0.99$ for dephasing errors up to $I_z^3$ and for single $I_{\pm}$ errors. Beyond this regime, the fidelity drops sharply, consistent with the correctable error set bounded by $l_{\mathrm{max}} \sim 2I/N$.

The small residual infidelity observed under pure dephasing originates from the finite overlap of the rotated logical states in the $I_z$ basis. Because the spin-coherent components have binomial distributions with nonzero tails, dephasing errors are corrected only asymptotically in $I$. Consequently, the logical fidelity approaches unity exponentially as the ensemble size increases.

\subsection{Realistic Gate Dynamics and Experimental Feasibility}

\begin{figure}[h]
    \centering
    \includegraphics[width=\linewidth]{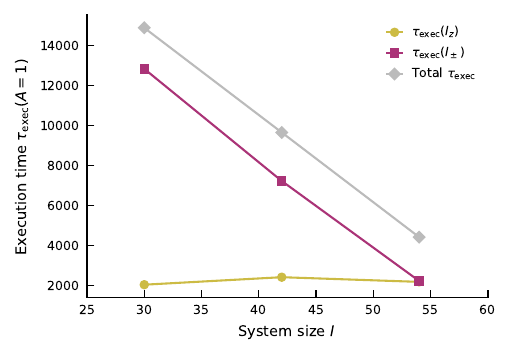}
    \caption{
    Simulation of a full error correction cycle under realistic experimental parameters. 
    The minimum $\tau_{\mathrm{exec}}$ required to maintain fidelity above \(99.8\%\) is shown as a function of system size $I$. 
    Larger ensembles enable faster correction through collective enhancement, especially in the \(I_\pm\)-correction stages.
    }
    \label{fig:realistic_plot}
\end{figure}

Unlike the idealized benchmark of Sec.~IV.A, we now simulate the full 6-Cat error-correction cycle including finite-duration gates generated by the effective Hamiltonian in Eq.~\eqref{eq:H_eff}. 
Figure~\ref{fig:realistic_plot} shows the minimum execution time $\tau_{\mathrm{exec}}$ required to maintain logical fidelity above $99.8\%$ (units $a=1$).

There is a clear trade-off between spectral selectivity and decoherence exposure: 
longer gates improve conditional addressing by enhancing frequency resolution, but increase $\tau_{\mathrm{exec}}$ and therefore the accumulated decoherence during the correction cycle.

As the ensemble size $I$ increases, flip-flop–based gates become faster due to collective enhancement. 
The matrix elements
\[
\langle I, M\pm 1 | I_\pm | I, M \rangle 
= \sqrt{I(I+1) - M(M\pm 1)}
\]
scale linearly with $I$ near the poles, accelerating $I_\pm$ transitions for large ensembles. 
As a result, the $I_\pm$-correction stage becomes progressively shorter with increasing $I$.

The $I_z$-correction stage also relies on flip-flop–based conditional operations and therefore benefits from the same collective speedup. 
However, the number of required operations grows with $I$, partially offsetting the faster gate times. 
Consequently, the $I_z$-correction duration remains approximately constant, as seen in Fig.~\ref{fig:realistic_plot}. 
For large ensembles, $I_z$-correction dominates the total cycle time, while $I_\pm$-correction becomes negligible.

Successful continuous error correction requires $\tau_{\mathrm{exec}} < \tau_{\max}$, where $\tau_{\max}$ is the maximum idle interval obtained from the idealized analysis in Fig.~\ref{fig:sim_ideal}. 
For moderate ensemble sizes ($I \sim 10^2$), we find $\tau_{\mathrm{exec}} \approx 1000/a$. 
As $I$ increases further, $\tau_{\mathrm{exec}}$ remains nearly constant while $\tau_{\max}$ grows with $I$, indicating improved viability of the code in the large-spin regime.

To assess experimental feasibility, we consider semiconductor quantum dots as a concrete realization of the central-spin architecture. Electron-mediated control of a nuclear ensemble has been demonstrated experimentally~\citep{Gangloff.2019,Appel.2025,Shofer.2024,Denning.2019}, and fast global nuclear rotations have been realized~\cite{Chekhovich2020}. These systems exhibit anisotropic hyperfine interactions and coherent driving, providing the ingredients required to engineer the gate primitives introduced in Sec.~III. In quantum dots, strain-induced quadrupolar interactions~\cite{Hogele2012a} and electron $g$-factor anisotropy~\cite{Shofer.2024} naturally generate noncollinear components of the hyperfine interaction, supplying the transverse coupling needed for conditional ensemble rotations.

Typical hyperfine couplings are $a \approx 500\,\mathrm{kHz}$~\citep{Appel.2025,Shofer.2024}. 
Because adjacent collective subspaces are separated by a frequency shift $\Delta\omega \sim a$, spectrally selective addressing requires pulse durations on the order of $t\sim 1/a \approx 2\,\mu$s. 
This timescale is comparable to reported central-spin dephasing times $T_2^* \approx 0.6\,\mu$s~\citep{Nguyen2023}. 
However, dynamical decoupling extends electronic coherence to $\sim 100\,\mu$s~\cite{Zaporski.2022}, and initialization of highly polarized collective states $\ket{I,\pm I}$~\cite{Appel.2025,Jackson.2022,Zaporski.2023} suppresses inhomogeneous broadening and mitigates the $T_2^*$ limitation. 
Nuclear coherence times reach $T_2 \approx 100\,\mathrm{ms}$~\citep{Dyte2025}, comfortably exceeding the full encoding, correction, and decoding times. 
Together, these parameters place both the gate execution time $\tau_{\mathrm{exec}}$ and the allowable correction interval $\tau_{\max}$ within experimentally accessible coherence windows, indicating that continuous implementation of the spin-$N$-Cat protocol is feasible in this platform.

Further improvements are possible using the two-qudit $I_z$-correction strategy introduced in Sec.~\ref{subsec:ErrorCorrection}, where the logical state is transferred between ensembles via a fixed sequence of fault-tolerant CNOT gates. 
In this approach, the number of $I_z$-correction operations becomes independent of $I$, and the CNOT duration inherits the collective enhancement of $I_\pm$ transitions. 
As a result, for large ensembles the total correction time can be substantially reduced relative to the single-ensemble protocol.
\section{Conclusion}  

We have introduced the spin-\( N \)-Cat code, a new class of quantum error-correcting codes that encode logical qubits in symmetric superpositions of spin-coherent states within a collective spin ensemble. By exploiting the modular phase-space structure of these superpositions, the code confines logical states and their correctable error spaces to well-separated subspaces labelled by \( M \bmod N/2 \). This partitioning enables syndrome-based error correction with minimal overhead and, in the large-\(I\) limit, uses nearly the entire Hilbert space.  

Crucially, our design enables the implementation of a universal, fault-tolerant, and bias-preserving gate set using only collective control and a single auxiliary spin. We have constructed these gates explicitly from experimentally accessible first-order angular momentum interactions, and demonstrated that the full QEC protocol can be realized in a central spin platform such as an electron-nuclear quantum dot with realistic parameters and gate times well within nuclear coherence limits.  

Numerical simulations confirm that the code satisfies the Knill–Laflamme conditions asymptotically, achieves high dephasing thresholds consistent with analytic bounds, and provides increasing robustness to excitation and decay as the code order \( N \) increases. Simulations with realistic gates further show that collective enhancement enables fast, high-fidelity correction cycles, making the protocol feasible even for moderate system sizes.  

Looking ahead, there are several directions to explore.  
A first step would be a hardware-level demonstration of the spin-\( N \)-Cat code in a suitable platform, such as a quantum dot with long-lived nuclear spins, which would serve as a strong proof of concept.  
Another avenue is to incorporate dissipative stabilization of the logical states, techniques that have proven effective in bosonic Cat codes~\citep{Mirrahimi.2014}, to further improve resilience against continuous noise.  
Finally, extending the approach to multiple spin-\( N \)-Cat logical qubits with native, bias-preserving gates could enable scalable multi-qubit architectures.  
Such couplings might be realized between different nuclear species within a single dot or by linking separate dots.  

Together, these results position the spin-\( N \)-Cat code as a scalable and physically grounded strategy for fault-tolerant quantum computation in large-spin systems, combining near-optimal use of the available Hilbert space with a hardware-compatible control scheme. 

\section{Acknowledgements}
We thank Simon Lieu, Apollonas S. Matsoukas-Roubeas, and Oscar Scholin for critical reading of our manuscript. We thank David Arvidsson-Shukur, Leon Zaporski, and Khadija Sarguroh for useful discussions. C.F. acknowledges funding from the German Academic Scholarship Foundation (Studienstiftung des deutschen Volkes), the Royal Society, and a Cavendish Laboratory scholarship. D.G. acknowledges a Royal Society University Research Fellowship and the QuantERA project MEEDGARD through EPSRC EP/Z000556/1. We acknowledge support from the University of Cambridge's High Performance Computing facilities.


\appendix

\section{ Knill--Laflamme Conditions} \label{sec:KL_appendix}
The Knill--Laflamme (KL) conditions define when a quantum code can correct a given error set \( \mathcal{E} \). They require
\begin{equation}
    P E_i^\dagger E_j P = \alpha_{ij} P \qquad \text{for all } E_i, E_j \in \mathcal{E},
\end{equation}
where \( P = |0\rangle_L \langle 0| + |1\rangle_L \langle 1| \) projects onto the logical subspace.

The spin-\(N\)-Cat code partitions the Hilbert space into \( N/2 \) orthogonal syndrome sectors, labeled by \( M \bmod N/2 \). 

\begin{figure}[htbp]
    \centering
    \includegraphics[width=\linewidth]{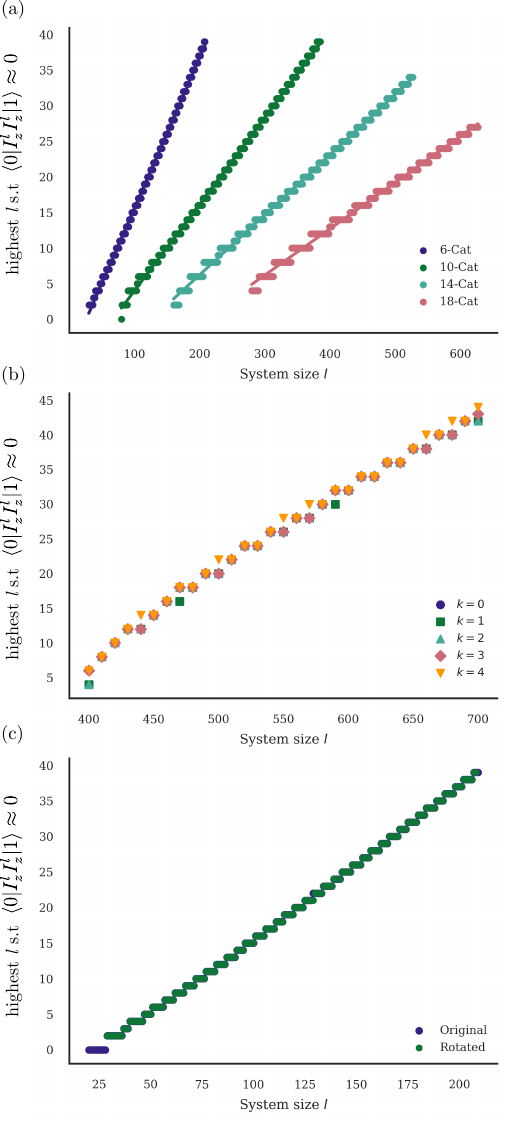}
    \caption{Numerical validation of the Knill--Laflamme conditions. (a) Number of dephasing errors satisfying the off-diagonal condition \( \langle 0 | I_z^m I_z^n | 1 \rangle_L \approx 0 \) for different \( N \). Each curve corresponds to a different Cat code order, showing linear growth with total spin \( I \), with slope \( \alpha \sim 1/N \). (b) Effect of increasing \( k \) (number of raising/lowering errors) on correctable dephasing errors ($N=21$). (c) Suppression of logical-state overlap in the rotated basis, confirming \( I_z \)-orthogonality after rotation.}
    \label{fig:KL_plot}
\end{figure}

Within each sector, dephasing-type errors \( I_z^m \) act non-trivially but remain confined. Since each sector has approximate dimension \( \sim 4I/N \), and both logical states must remain distinguishable under these errors, the number of correctable dephasing errors is bounded by
\[
l_{\max} \sim \frac{2I}{N}.
\]
Numerically, we observe the overlap
\[
\langle 0 | I_z^m I_z^n | 1 \rangle_L \approx 0 \quad \text{for all } m,n \leq l = \alpha I + b,
\]
with \( \alpha \approx 2/N \), in agreement with this bound. See Fig.~\ref{fig:KL_plot} (top panel). The small \(N\)-dependent offset \(b\) is expected: the spin-\(N\)-Cat code is designed for physical implementability using only first-order collective interactions, rather than to saturate the Hilbert space bound. The resulting geometric structure and symmetry constraints limit the available degrees of freedom, introducing a tradeoff between maximal correctability and experimental accessibility. In the large-$I$ limit the offset becomes negligible, and the modular subspaces are nearly perfectly partitioned into orthogonal two-dimensional blocks.

In the presence of excitation-type errors, the logical states are mapped to different syndrome sectors. The KL condition then requires
\begin{equation}
\langle 0 | I_\pm^k I_z^m I_z^n I_\pm^k | 1 \rangle_L \approx 0 \quad \text{for } k \leq \frac{N - 2}{4}, \quad m,n \leq l.
\end{equation}
This is satisfied because \( I_z \) remains diagonal and localized within each syndrome sector, and the effect of \( I_\pm^k \) on the logical state shape is small for \( k \ll I \).

Geometrically, applying \( k \) raising or lowering operators shifts the logical state to polar angle
\[
\theta_k = \cos^{-1}(k/I),
\]
and reduces its effective equatorial radius to
\[
r_k = I \sin\theta_k \approx I - \frac{k^2}{2I} + \mathcal{O}\left( \frac{k^4}{I^3} \right).
\]
This reduction vanishes in the large-\( I \) limit, so the dephasing tolerance is unaffected at leading order. See Fig.~\ref{fig:KL_plot} (bottom left).

The diagonal condition
\[
\langle 0 | E^\dagger E | 0 \rangle_L \approx \langle 1 | E^\dagger E | 1 \rangle_L
\]
also holds for mixed errors due to the symmetry of the spin-coherent components. Each error affects the logical components uniformly, preserving expectation values.

The main text (Sec. III) introduces a rotated codespace in which logical states acquire distinct \( I_z \)-distributions, making them distinguishable under conditional gates. Numerically, we confirm that overlaps \( \langle 0 | I_z^m I_z^n | 1 \rangle_L \) vanish in this basis for increasing \( I \), consistent with asymptotic KL satisfaction. See Fig.~\ref{fig:KL_plot} (bottom right).

Together, these results confirm that the spin-\( N \)-Cat code satisfies the KL conditions for all errors in the set \( \mathcal{E}_{k,l} \), including mixed compositions of excitation and dephasing, in the large-\( I \) regime. This establishes the formal error-correcting capability of the code structure introduced in Section~II.

\subsection*{Numerical Validation Using Chebyshev Polynomials}

To assess the correctability of dephasing errors, we verify the off-diagonal Knill–Laflamme condition
\[
\langle 0 | I_z^m | 1 \rangle \approx 0
\]
for all \( m \leq l \). Rather than testing powers of \( I_z \) directly, which becomes numerically unstable for large \( m \), we compute
\[
\langle 0 | T_m(I_z) | 1 \rangle,
\]
where \( T_m(x) \) is the Chebyshev polynomial of the first kind. These form an orthogonal basis for functions on the interval \( x \in [-1,1] \), and each \( T_m(x) \) is a degree-\( m \) polynomial in \( x \). We first rescale \( I_z \) by \( I \) to ensure its spectrum lies within \( [-1, 1] \), then construct the operator \( T_m(I_z) \) by diagonalization.

Because \( T_0, \ldots, T_m \) span the space of all degree-\( m \) polynomials \citep{Trefethen.2019}, the condition
\[
\langle 0 | T_j(I_z) | 1 \rangle \approx 0 \quad \text{for all } j \leq m
\]
implies
\[
\langle 0 | f(I_z) | 1 \rangle \approx 0
\]
for all polynomial functions \( f(I_z) \) of degree \( \leq m \). In particular, this ensures that all matrix elements \( \langle 0 | I_z^j | 1 \rangle \approx 0 \) for \( j \leq m \), thus validating the Knill–Laflamme condition for dephasing-type errors up to order \( m \).

This approach allows stable numerical estimation of the correctable error order \( l \), even for large \( I \), and is used throughout to quantify the dephasing threshold shown in Fig.~\ref{fig:KL_plot}.
In our implementation, we apply the Chebyshev operator \( T_m(I_z) \) to both logical states before computing the overlap, effectively checking matrix elements of the form \( \langle 0 | T_m(I_z)^2 | 1 \rangle \). This ensures suppression of off-diagonal terms involving composed dephasing errors of total degree up to \( 2m \), including interference between different error orders.

\section{Logical Error Rate of the Two-Level Dicke Encoding}
\label{app:t_dicke}

In this appendix we derive an explicit expression for the characteristic
coherence time \(t_{\mathrm{Dicke}}(\eta)\) of an unprotected two-level
encoding under collective noise. This serves as the reference for the
improvement factor \(\mathcal{R}(N,\eta)=
\tau_{\mathrm{max}}(N,\eta)/t_{\mathrm{Dicke}}(\eta)\)
used in the main text.

\subsection{Model}

The Dicke encoding is defined by the states
\(|0\rangle = |I,-I\rangle\) and \(|1\rangle = |I,-I+1\rangle\).
The ensemble experiences collective dephasing and collective
raising/lowering at rates \(\gamma_z\), \(\gamma_+\), and \(\gamma_-\),
normalized such that
\(\gamma_z+\gamma_++\gamma_-=1\).
We parametrize the relative noise bias as
\[
\eta=\frac{\gamma_z}{\gamma_++\gamma_-},\qquad
\gamma_z=\frac{\eta}{1+\eta},\qquad
\gamma_++\gamma_-=\frac{1}{1+\eta}.
\]

The logical error rate is quantified by the
average gate infidelity \(r_{\mathrm{avg}}(t)=1-F_{\mathrm{avg}}(t)\)
of the channel obtained after time~\(t\) under the collective noise.
A correction cycle is considered successful if
\(F_{\mathrm{avg}}(t)\ge0.999\),
corresponding to a target per-cycle error
\(\varepsilon=10^{-3}\).

\subsection{Effective logical dynamics}

\paragraph{Collective dephasing.}
For \(L_z=\sqrt{\gamma_z}\,I_z\) the master equation reads
\[
\dot\rho=\gamma_z\left(I_z\rho I_z-\tfrac12\{I_z^2,\rho\}\right).
\]
Within the subspace spanned by \(|0\rangle,|1\rangle\),
the collective operator acts as
\(I_z=(-I+\tfrac12)\openone_L-\tfrac12 Z_L\),
so that the dissipator reduces to
\(\dot\rho_L=(\gamma_z/4)(Z_L\rho_LZ_L-\rho_L)\).
Hence the off-diagonal elements decay as
\(\rho_{01}(t)=e^{-(\gamma_z/2)t}\rho_{01}(0)\),
i.e.~the logical qubit undergoes a dephasing channel with coherence factor
\(\lambda(t)=e^{-(\gamma_z/2)t}\).
The corresponding average infidelity contribution is \citep{nielsen2010quantum}
\begin{equation}
r_z(t)=\tfrac{1}{3}\!\left[1-e^{-(\gamma_z/2)t}\right].
\label{eq:rz}
\end{equation}

\paragraph{Collective raising and lowering.}
For \(L_\pm=\sqrt{\gamma_\pm}\,I_\pm\), a single jump produces a logical
bit flip or leakage:
\[
I_-|1\rangle\propto|0\rangle,\qquad
I_+|0\rangle\propto|1\rangle,\qquad
I_+|1\rangle\propto|I,-I+2\rangle.
\]
Averaging over logical inputs gives an estimated first-jump rate
\begin{equation}
r_{\mathrm{jump}} \simeq 2I\,(\gamma_+ + \gamma_-),
\label{eq:rpm}
\end{equation}
reflecting the collective enhancement of ladder transitions.
Near the south pole ($M\simeq -I$), the matrix elements
$I_\pm|I,M\rangle \propto \sqrt{2I}$ imply transition probabilities
scaling as $2I$, since any of the $2I$ spins can flip within the
symmetric manifold.  The logical error rate under ladder noise thus grows
linearly with $I$, consistent with the expected collective coupling
scaling.

\subsection{Total error and time}

Combining Eqs.~\eqref{eq:rz}--\eqref{eq:rpm} gives the total
average infidelity after an uncorrected evolution time~\(t\),
\begin{equation}
r_{\mathrm{avg}}(t)
=\frac{1-e^{-(\gamma_z/2)t}}{3}
+2I(\gamma_++\gamma_-)\,t.
\label{eq:ravg}
\end{equation}
The characteristic time \(t_{\mathrm{Dicke}}(\eta)\) is obtained by solving
\(r_{\mathrm{avg}}(t)=\varepsilon\) for \(t\).
For the small target error \(\varepsilon\ll1\), it is sufficient to
expand the dephasing contribution to first order in time,
\(
1-e^{-(\gamma_z/2)t}\approx (\gamma_z/2)t
\).
Equation~\eqref{eq:ravg} then reduces to the linear form
\[
r_{\mathrm{avg}}(t)
\approx
\left(
\frac{\gamma_z}{6}
+
2I(\gamma_++\gamma_-)
\right)t .
\]
Solving for \(t\) gives the characteristic Dicke coherence time
\begin{equation}
t_{\mathrm{Dicke}}(\eta)
\approx
\frac{\varepsilon}{
\frac{\gamma_z}{6}
+
2I(\gamma_++\gamma_-)
}.
\label{eq:t_dicke}
\end{equation}

\section{Physical Gates}\label{sec:Phys_Gates_Appendix}

The central spin system allows for indirect control of the collective nuclear ensemble via interactions mediated by a single electronic spin. All gate operations used in the code -- conditional ensemble rotations, conditional qubit rotations, and flip-flop transitions -- can be derived from a common system Hamiltonian. By applying time-dependent driving fields and pulse sequences, we isolate different effective interactions from this underlying model.

We begin with the undriven Hamiltonian:
\begin{equation}
    H = \omega_e S_z + \omega_n I_z + aS_z I_z + a_{nc} S_z I_x,
\end{equation}
where \( \omega_e \) and \( \omega_n \) are the Zeeman energies of the central spin and the ensemble, respectively. The collinear interaction \( A S_z I_z \) produces an ensemble-dependent shift in the qubit frequency, while the noncollinear term \( A_{nc} S_z I_x \) enables exchange of excitations between the two systems. Depending on the applied control, each term in this Hamiltonian can be enhanced, suppressed, or filtered to yield the desired gate.

When a microwave drive is applied to the central spin, the Hamiltonian becomes time-dependent:
\begin{equation}
\begin{aligned}
    H(t) = &\omega_e S_z + \omega_n I_z + a S_z I_z + a_{nc} S_z I_x \\&+2\Omega \cos(\omega t + \phi) S_x,
\end{aligned}
\end{equation}
where \( \Omega \) is the Rabi amplitude, \( \omega \) the drive frequency, and \( \phi \) the drive phase. We extend the system to include a multi-tone drive:
\begin{equation}
\begin{aligned}
    H(t) &= \omega_e S_z + \omega_n I_z + a S_z I_z + a_{nc} S_z I_y \\
    &\quad+2\Omega \sum_k \cos(\omega_k t + \varphi_k)\, S_x,
\end{aligned}
\end{equation}
with \( \omega_k \) chosen depending on the gate. Moving to a frame rotating at the electron frequency via \( U(t) = e^{i \omega_e S_z t} \), we obtain oscillating terms at both \( \omega_k - \omega_e \) and \( \omega_k + \omega_e \). The rotating wave approximation (RWA) discards the latter, which oscillate rapidly at frequencies \( \omega_k + \omega_e \), and thus average out over the timescale of the dynamics provided \( \Omega \ll \omega_k + \omega_e \)~\cite{CohenTannoudji1992,ScullyZubairy1997}. The resulting interaction-frame Hamiltonian becomes:
\begin{equation}
\begin{aligned}
    H_{\text{RWA}} &= \omega_n I_z + a S_z I_z + a_{nc} S_z I_y \\
    &\quad+ \Omega \sum_k \left[ \cos(\Delta_k t + \varphi_k)\, S_x + \sin(\Delta_k t + \varphi_k)\, S_y \right],
\end{aligned}
\label{eq:H_RWA}
\end{equation}
where \( \Delta_k = \omega_k - \omega_e \) is the detuning of each drive tone from the qubit frequency. Although the detuning \( \Delta_k \) varies depending on the gate, all \( \omega_k \) remain close to \( \omega_e \), so the condition \( \Omega \ll \omega_k + \omega_e \) remains valid across all cases.

In the following, we show how tailored pulse sequences and resonant drives give rise to the physical operations required for encoding, logical gates, and error correction.

\subsection{Engineered Ensemble Rotations via Pulse Sequences}

\begin{figure}
    \centering
    \includegraphics[width=1.0\linewidth]{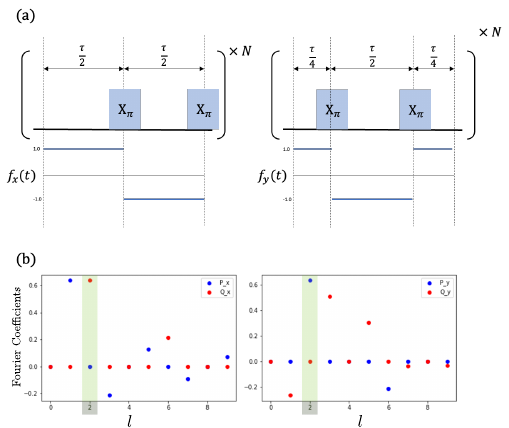}
    \caption{Engineered ensemble rotations via pulse sequences. 
(a) Pulse sequences consisting of $\pi$ pulses on the central spin (blue boxes) separated by free evolution intervals, generating square-wave modulation functions $f_x(t)$ and $f_y(t)$ that modulate the interaction Hamiltonian. 
(b) Fourier coefficients of the modulation functions from pulse optimization; the highlighted $\ell=2$ harmonic provides the resonant component used to engineer effective interactions $H_x \approx 0.63\, S_z I_x$ and $H_y \approx 0.63\, S_z I_y$.}
    \label{fig:pulse_sequence}
\end{figure}

To implement ensemble rotations conditioned on the central spin state, we enhance the weak \( S_z I_x \) interaction using periodic pulse sequences. These sequences modulate the sign of \( S_z \) in time, selectively amplifying desired terms in the interaction Hamiltonian while suppressing others. We apply a pattern of central spin \( \pi \)- and \( \pi/2 \)-pulses interleaved with free evolution, as shown in Fig.~\ref{fig:pulse_sequence}a. This modifies the Hamiltonian to:
\begin{equation}
    H(t) = \omega_n I_z + f_{x/y}(t)\, S_z \left( \omega_e + a I_z + a_{nc} I_x \right),
\end{equation}
where \( f_{x/y}(t) \) is a square-wave modulation function, differing by a phase shift of \( \tau/4 \) between \( x \)- and \( y \)-rotations.

This function can be expanded in a Fourier series:
\begin{equation}
\begin{aligned}
    f_{x/y}(t) = &\sum_{\ell=0}^\infty \left[ P_\ell^{(x/y)} \cos\left( \omega_\ell t \right) + Q_\ell^{(x/y)} \sin\left( \omega_\ell t \right) \right], \\ \omega_\ell = &\frac{\pi \ell}{\tau},
\end{aligned}
\end{equation}
with coefficients:
\begin{equation}
\begin{aligned}
    P_\ell^{(x/y)} &= \frac{1}{\tau} \int_0^{2\tau} f_{x/y}(t) \cos\left( \frac{\pi \ell t}{\tau} \right) dt, \\
    Q_\ell^{(x/y)} &= \frac{1}{\tau} \int_0^{2\tau} f_{x/y}(t) \sin\left( \frac{\pi \ell t}{\tau} \right) dt.
\end{aligned}
\end{equation}

Transforming to a rotating frame with \( \omega_n I_z \), the nuclear operator becomes:
\begin{equation}
    I_x \rightarrow \frac{1}{2} \left( e^{i \omega_n t} I_+ + e^{-i \omega_n t} I_- \right).
\end{equation}

To isolate the resonant component, we choose the pulse period such that \( \omega_\ell = \omega_n \), i.e., \( \tau = \pi \ell / \omega_n \). For \( \ell = 2 \), and neglecting fast-oscillating terms, we obtain:
\begin{equation}
\begin{aligned}
    H_{\text{eff}} = &\frac{1}{2} \Big[ \left( P_2^{(x/y)} + i Q_2^{(x/y)} \right) S_z I_+ 
    \\&+ \left( P_2^{(x/y)} - i Q_2^{(x/y)} \right) S_z I_- \Big].
\end{aligned}
\end{equation}

From pulse optimization (see Fig.~\ref{fig:pulse_sequence}b), we find:
\[
P_2^{(x)} = Q_2^{(y)} = 0, \quad Q_2^{(x)} = P_2^{(y)} \approx 0.63,
\]
leading to effective interactions:
\begin{equation}
    H_x \approx 0.63\, S_z I_x, \qquad H_y \approx 0.63\, S_z I_y.
\end{equation}

This gate benefits from a large nuclear frequency \( \omega_n \), which ensures the rotating wave approximation is valid and off-resonant Fourier components are suppressed.

\subsection{Conditional Qubit Rotation via Multi-Tone Resonant Drive}

To perform rotations of the central spin conditioned on the state of the ensemble, we operate Eq.~\eqref{eq:H_RWA} in the near-resonant regime. The hyperfine shift \( a I_z \) modifies the qubit resonance depending on the collective state \( \ket{I,n} \), resulting in an effective detuning \( \delta' = \delta + a n \). Driving at \( \omega = \omega_e + a n \) brings the electron into resonance with that subspace.

The maximum probability of a spin flip is given by:
\[
P_{\text{flip}}^{\text{max}} = \frac{\Omega^2}{\Omega^2 + \delta'^2},
\]
which quantifies the suppression of off-resonant transitions. A smaller Rabi amplitude \( \Omega \) improves selectivity, though it increases the gate duration and susceptibility to decoherence.

To target multiple ensemble states simultaneously, we apply a multi-tone drive of the form:
\begin{equation}
    \Omega(t) = \sum_k \Omega_k \cos((\omega_e + a k)t + \theta_k),
\end{equation}
where each component is resonant with a different \( \ket{I,k} \) subspace. The phases \( \theta_k \) compensate for hyperfine-induced phase accumulation during the gate:
\[
\theta_k = \left( \frac{\psi}{\Omega_k} \cdot A k \right) \bmod 2\pi,
\]
with \( \psi \) the desired rotation angle.

In the rotating frame, the drive term becomes:
\begin{equation}
\begin{aligned}
H_{\text{RF}} =& \omega_n I_z + a I_z S_z + a_{nc} S_z I_x 
 \\&+ \sum_k \Omega_k\Big[ \cos(-a k t + \theta_k) S_x \\&\quad- \sin(-a k t + \theta_k) S_y \Big].
\end{aligned}
\end{equation}

This enables simultaneous, state-selective qubit rotations using a single multi-frequency pulse. High-fidelity operation requires \( \Omega \) to be small enough to suppress off-resonant excitations. Additionally, working in the regime \( A_{nc} \ll A \) and \( \omega_n \gg a_{nc} \) ensures that the noncollinear interaction remains off-resonant and has negligible effect on the rotation. \\
Alternatively, the noncollinear interaction \( a_{\text{nc}} S_z I_x \) can be dynamically refocused using a decoupling sequence applied to the ensemble. We use an XY8 sequence of the form:
\[
\begin{array}{cccccccc}
X & \xrightarrow{\tau} & Y & \xrightarrow{\tau} & X & \xrightarrow{\tau} & Y & \xrightarrow{\tau} \\
Y & \xrightarrow{\tau} & X & \xrightarrow{\tau} & Y & \xrightarrow{\tau} & X &
\end{array}
\]
Each \( \pi \)-pulse flips the sign of the \( I_x \) operator, effectively averaging the noncollinear term to zero over the full cycle. In contrast, the collinear hyperfine term \( a S_z I_z \) remains unaffected, as \( I_z \) commutes with all pulses in the sequence. This selective decoupling isolates the desired \( S_z I_z \) interaction while suppressing unwanted terms.

During a conditional qubit rotation, the XY8 sequence modifies the system’s resonance condition dynamically, and this evolution must be accounted for to maintain selective addressing. Each \( \pi \)-pulse maps the collective state \( \ket{I, m} \) to \( \ket{I, -m} \), effectively flipping the sign of the quantization axis. As a result, the resonance condition shifts dynamically during the sequence. During the sequence, the resonance condition for a given state evolves as: 
\begin{equation}
    \omega_k(t) = (-1)^n a k t + \delta\omega(t),
\end{equation}
where $n(t) = \left\lfloor \frac{t}{\tau} \right\rfloor$ counts the number of $\pi$-pulses applied and $\delta\omega(t)=(-1)^{n(t)} \cdot 2 \tau a k \cdot \left\lfloor \frac{n(t)}{2} \right\rfloor$ accounts for the accumulated phase shifts due to the hyperfine phases.

In addition to toggling the noncollinear term, the ensemble undergoes rotation under the combined hyperfine and nuclear Zeeman interaction during the gate. The resulting accumulated phase can either be accounted for in subsequent gate calibrations, or the total gate duration can be chosen such that this phase becomes a multiple of \( 2\pi \), effectively refocusing it at the end of the sequence.

\subsection{Flip-Flop Gate in the Sideband Limit}

We implement selective flip-flop transitions between the central spin and the ensemble by operating Eq.~\eqref{eq:H_RWA} in a regime where direct spin flips are suppressed and second-order processes become resonant. These transitions are mediated by the noncollinear interaction \( a_{nc} S_z I_y \) in the presence of a weak, off-resonant drive.

To isolate the slow dynamics, we perform a perturbative unitary transformation to eliminate the fast non-commuting term. We split the Hamiltonian as
\[
H(t) = H_0 + H_0'(t) + V,
\]
with
\[
H_0 = \omega_n I_z, \qquad V = a_{nc} S_z I_y,
\]
and
\[
\begin{aligned}
   H_0'(t) &= a S_z I_z \\&\quad+ \Omega \sum_k \left[ \cos(\Delta_k t + \varphi_k)\, S_x + \sin(\Delta_k t + \varphi_k)\, S_y \right]. 
\end{aligned}
\]

We construct the generator \( G_1 \) satisfying \( [H_0, G_1] = -V \), which gives
\[
G_1 = -i \frac{a_{nc}}{\omega_n} S_z I_x.
\]

Expanding the effective Hamiltonian to second order yields
\[
H_{\text{eff}} = H_0 + H_0'(t) + \frac{1}{2} [G_1, H_0'(t)].
\]

The commutator yields two contributions. First, from the drive term:
\begin{equation*}
    \begin{aligned}
        [G_1, H_{\text{drive}}(t)] &= \Big[ -i \frac{a_{nc}}{\omega_n} S_z I_x, \\
&\quad\Omega \sum_k \left( \cos(\Delta_k t + \varphi_k)\, S_x + \sin(\Delta_k t + \varphi_k)\, S_y \right) \Big] \nonumber \\
&= \frac{a_{nc} \Omega}{\omega_n} \sum_k \Big[ \cos(\Delta_k t + \varphi_k)\, S_y \\
&\quad- \sin(\Delta_k t + \varphi_k)\, S_x \Big] I_x.
    \end{aligned}
\end{equation*}

Second, the collinear interaction contributes
\[
[G_1, a S_z I_z] = -i \frac{a a_{nc}}{\omega_n} [S_z I_x, S_z I_z] = -i \frac{a a_{nc}}{\omega_n} S_z I_y.
\]
This term does not match the resonance condition for any transition and can therefore be neglected.

The resulting effective Hamiltonian becomes:
\begin{equation*}
\begin{aligned}
H_{\text{eff}} =\ &\omega_n I_z + a S_z I_z \\&\quad+ \Omega \sum_k \left[ \cos(\Delta_k t + \varphi_k)\, S_x + \sin(\Delta_k t + \varphi_k)\, S_y \right] \\
&+ \frac{a_{nc} \Omega}{\omega_n} \sum_k \left[ \cos(\Delta_k t + \varphi_k)\, S_y - \sin(\Delta_k t + \varphi_k)\, S_x \right] I_x.
\end{aligned}
\end{equation*}

The last term describes a second-order interaction \( \sim S_{x,y} I_x \) that becomes resonant when the drive frequency matches the energy difference between the states \( \ket{\uparrow, M} \) and \( \ket{\downarrow, M \pm 1} \). This difference is set by the diagonal part of the Hamiltonian, \( \omega_e S_z + \omega_n I_z + a S_z I_z \), which gives
\[
\Delta E = \omega_e \pm \omega_n + a(M \pm \tfrac{1}{2}),
\]
with the signs depending on whether the nuclear spin flips up or down and whether the central spin flips up or down. The drive frequency must be tuned to match this energy difference. Both flip-flop (\( S_- I_+ \), \( S_+ I_- \)) and flip-flip (\( S_+ I_+ \), \( S_- I_- \)) transitions can in principle be resonantly driven, depending on the detuning.

The effective interaction strength is
\[
a_{nc}' = \frac{\Omega a_{nc}}{\omega_n},
\]
and the gate time for the transition \( \ket{I,M} \to \ket{I,M+1} \) is
\[
t_M = \frac{2\pi}{a_{nc}' \sqrt{I(I+1) - M(M+1)}}.
\]

The Rabi frequency \( \Omega \) must remain small compared to the detuning \( \delta \) to ensure the validity of the sideband approximation and suppress off-resonant excitations. The flip-flop gate benefits from a strong noncollinear interaction \( a_{nc} \), which enhances the effective coupling strength \( a_{nc}' = \Omega a_{nc} / \omega_n \) and reduces gate time. A moderate nuclear Zeeman energy \( \omega_n \) improves selectivity by increasing the spectral separation between transitions. This contrasts with the conditional qubit rotation gate, which requires \( a_{nc} \ll a \) and large \( \omega_n \) to achieve subspace resolution. Together, the two gates constrain the optimal operating regime of the system, reflecting a trade-off between speed, fidelity, and addressability.

\section{Error Correction}
\subsection{\( I_+/I_- \)-Error Correction}

To simulate the decay and pumping error correction described in the main text, we must carefully account for all relative phase shifts that arise during the correction sequence. The procedure begins by engineering a selective flip-flop transition on the \( M \bmod 3 = 1 \) subspace. During this gate, only this subspace becomes entangled with the excited state of the electron, while the rest of the ensemble remains associated with the ground state.

Due to the hyperfine interaction, these two parts of the Hilbert space precess at different rates: the \( M \bmod 3 = 1 \) subspace evolves at frequency \( \omega_n - \frac{a}{2} \), while the rest evolves at \( \omega_n + \frac{a}{2} \). To realign the relative phase between these subspaces, we insert a free evolution period after the gate. The required duration is
\[
t_{\text{free}} = \left\lceil \frac{a t_{\text{flip}}}{4\pi} \right\rceil \cdot \frac{4\pi}{a} - t_{\text{flip}},
\]
which ensures that the entire Hilbert space refocuses before reinitializing the electron.

In addition, we must correct for the global phase acquired by the ensemble due to nuclear Zeeman precession. This introduces an overall rotation of
\[
\theta = \left[ (\omega_n)(t_{\text{flip}} + t_{\text{free}}) \right] \bmod 2\pi,
\]
which we account for by applying a corrective phase shift to the ensemble.

An analogous procedure is used for correcting \( I_- \)-type errors via a flip-flip transition, with the same considerations for phase alignment and ensemble precession.

\subsection{Dephasing Error Correction}

As in the decay/pumping correction scheme, the dephasing correction sequence must carefully account for relative phase accumulation between ensemble states. Specifically, the coherent superposition states \( \ket{+l} + e^{i \Delta \omega} \ket{-l} \) acquire different phases depending on the path taken through the correction circuit. When repumping the electron, we combine the states
\[
(\ket{l} + e^{i \Delta \omega} \ket{-l}) \otimes \ket{\uparrow}, \quad \text{and} \quad (\ket{l} + e^{i \Delta \omega'} \ket{-l}) \otimes \ket{\downarrow},
\]
so it is crucial that \( \Delta \omega = \Delta \omega' \) to avoid destructive interference. This condition must be enforced by choosing the flip-flop gate time for the \( (l-1 \rightarrow l) \) transition appropriately.

Since the full correction procedure consists of multiple sequential flip-flop operations, each contributes to the total phase accumulation. Let \( T \) denote the total duration of all prior gates before the current transition. Then, the accumulated phase differences are:
\begin{equation}
\begin{aligned}
\Delta \omega &= (T + t_{l-1 \rightarrow l}) \cdot 2l (\omega_n + \tfrac{1}{2}), \\\\
\Delta \omega' &= T \cdot 2(l-1)(\omega_n + \tfrac{1}{2}) + t_{l-1 \rightarrow l} \cdot 2l(\omega_n - \tfrac{1}{2}).
\end{aligned}
\end{equation}
Imposing the condition \( \Delta \omega = \Delta \omega' + 2\pi k \), we solve for the gate duration:
\begin{equation}
t_{l-1 \rightarrow l} = \frac{k\pi + T(\omega_n + \tfrac{1}{2})}{l},
\end{equation}
and adjust the Rabi amplitude \( \Omega \) accordingly to achieve this timing.

After the final flip-flop operation, a residual phase offset remains between the most polarized states \( \ket{-N} \) and \( \ket{+N} \). This offset is given by:
\[
\theta = 2N (\omega_n + \tfrac{1}{2}) T_{\text{tot}} \bmod 2\pi,
\]
where \( T_{\text{tot}} \) is the total duration of the correction sequence. To restore coherence between these states, we apply a final free evolution of duration:
\[
t_{\text{refocus}} = \frac{2\pi - \theta}{2N(\omega_n + \tfrac{1}{2})}.
\]

\section{Logical Gates} \label{Appendix_gates}
\subsection{CNOT – ensemble control qubit}
We define the rotated error subspaces \(\mathcal{S}_j^{(k,\ell)}\) as
\[
\mathcal{S}_j^{(k,\ell)} := \text{span} \left\{ e^{-i\frac{\pi}{2} I_y} E \ket{j}_L \mid E \in \mathcal{E}_{k,\ell} \right\}, \quad j = 0,1,
\]
which represent the images of the logical codewords under correctable collective noise, transformed into the rotated codespace (see Fig.~\ref{fig:physical_implementation}d).  

The support set \(\Gamma^{(k,\ell)}\) contains all collective spin projections \(M\) that occur in \(\mathcal{S}^{(k,\ell)}_0\):
\[
\Gamma^{(k,\ell)} := \{ M \mid \langle I,M | \psi \rangle \neq 0 \ \text{for some} \ |\psi\rangle \in \mathcal{S}^{(k,\ell)}_0 \}.
\]
In practice, for the 6-Cat code we use here, \(\Gamma\) is the union of the ranges
\begin{align*}
n &\in [I - N/6, I] \quad\text{and}\\\quad n &\in \left[ \left\lceil\frac{2N+1-p^{-2}}{1+p^{-2}}\right\rceil - \frac{m}{2}, \ \left\lceil\frac{2N+1-p^{-2}}{1+p^{-2}}\right\rceil + \frac{m}{2} \right],
\end{align*}
with \(p = \sqrt{3}\), which correspond to the most polarized states and the set of states within \(m/2\) of the central support peak.

After the \(\pi/2\) rotation about the \(y\)-axis, a central spin rotation conditional on \(\Gamma^{(k,\ell)}\) implements a CNOT with the ensemble as control (Fig.~\ref{fig:gates_sq}a). The conditional rotation acts on \(|0\rangle_L\) and all of its correctable error subspaces, while leaving \(|1\rangle_L\) and its associated error subspaces unchanged, thereby satisfying criterion~(ii).

In the physical implementation, the \(\pi/2\) rotation about \(y\) maps the logical states into an \(I_z\)-distinguishable basis. A \(\Pi(\pi,0,\Gamma)\) gate then rotates the electron conditional on the ensemble being in the \(|0\rangle_L\) subspace as defined above. An ensemble rotation \(I_y(-\pi/2)\) returns the system to the original codespace. Because \(\exp\{-\frac{i}{2}\pi X\} = -iX\), the resulting two-qubit operation is
\[
    \begin{bmatrix}
    1&0&0&0\\
    0&1&0&0\\
    0&0&-i&0\\
    0&0&0&-i
\end{bmatrix},
\]
introducing only a global phase on the electron spin, which must be tracked in subsequent operations.

Fault tolerance is ensured because the rotated subspaces are well separated in the \(I_z\)-basis, so the conditional operation does not couple different syndrome spaces. Bias preservation follows because the ensemble undergoes a \(\pi/2\) rotation about \(y\) and is returned by a \(-\pi/2\) rotation, leaving its polarization, and thus \(M \bmod \tfrac{N}{2}\), unchanged. This guarantees that a dephasing-type error remains dephasing-type and is not converted to an \(I_\pm\)-type error.

\subsection{CNOT - central spin control qubit}
If the central spin is the control qubit, the collinear interaction $aS_zI_z$ can be used to introduce different precession times on the ensemble spins. Waiting a time $t=\frac{\pi}{a}$ yields a logical bit flip on the ensemble if the electron is in the up state compared to the down state. \\
If a raising or lowering error occured, the average angular momentum projection is not zero anymore, but $\pm A$. This means that during the waiting time the electron also accumulates a phase $\pi$.  A $\Pi(\pi,0,\sum{n|n\bmod{3}=\{1,2\}})\cdot\Pi(\pi,\pi,\sum{n|n\bmod{3}=\{1,2\}})$ gate corrects this phase to ensure that the gate is fault-tolerant (Here we decompose the Z-gate as $Z=R_x(\pi)R_y(\pi)$). 
\subsection{Hadamard}

The Hadamard gate starts with a CNOT gate with the ensemble qubit as control. After this step, the \(\ket{1}_L\) logical state is entangled with the electron \(\ket{\uparrow}\) state and acquires a different precession frequency from the \(\ket{0}_L\) state due to the collinear interaction \(a S_z I_z\).  

The system then evolves freely for a time \(t = \frac{\pi}{3A}\), at which point the two ensemble states overlap in the collective basis but remain entangled with different electron states. A single-qubit unitary \(U\) is applied to the electron to cancel the relative phase accumulated between the logical components during the conditional evolution. This is followed by a second period of free evolution for the same duration and a final CNOT (ensemble control) that disentangles the electron from the ensemble, completing the Hadamard operation.

The full evolution can be represented as:
\begin{align*}
\ket{+}\ket{\downarrow} 
&\xrightarrow{i\mathrm{CNOT}} \ket{0}\ket{\downarrow} + i\ket{1}\ket{\uparrow} \\
&\xrightarrow{t} \ket{0}\left[\ket{\downarrow} + e^{i\theta}\ket{\downarrow}\right] \\
&\xrightarrow{U} \ket{0}\ket{\downarrow} \\
&\xrightarrow{t} \ket{0}\ket{\downarrow} \\
&\xrightarrow{i\mathrm{CNOT}} \ket{0}\ket{\downarrow} \\
\\
\ket{0}\ket{\downarrow} 
&\xrightarrow{i\mathrm{CNOT}} \ket{0}\ket{\downarrow} \\
&\xrightarrow{t} \ket{0}\ket{\downarrow} \\
&\xrightarrow{U} \ket{0}\left[\ket{\downarrow} + e^{-i\theta}\ket{\downarrow}\right] \\
&\xrightarrow{t} \ket{0}\ket{\downarrow} + e^{-i\pi/2} \ket{1}\ket{\uparrow} \\
&\xrightarrow{i\mathrm{CNOT}} \ket{\downarrow}\left[\ket{0} + \ket{1}\right]
\end{align*}

Here \(\theta = \omega_e t\) in the absence of errors. If a raising or lowering error has occurred, the ensemble acquires a nonzero average magnetization, shifting the precession frequency and modifying the phase to \(\theta = \omega_e t \pm A t\).  
To correct for this, we choose \(U\) to satisfy:
\[
U\ket{\downarrow} \overset{!}{=} \ket{\downarrow} + e^{-\theta}\ket{\uparrow}, \quad
U\left(\ket{\downarrow} + e^{i\theta}\ket{\uparrow}\right) \overset{!}{=} \ket{\downarrow}.
\]
A general single-qubit unitary can be written \cite{nielsen2010quantum} as:
\[
    U=\begin{bmatrix}
        e^{i(\alpha-\beta/2-\delta/2)}\cos{\gamma/2} & -e^{i(\alpha-\beta/2+\delta/2)}\sin{\gamma/2} \\
        e^{i(\alpha+\beta/2-\delta/2)}\sin{\gamma/2} & -e^{i(\alpha+\beta/2+\delta/2)}\cos{\gamma/2}
    \end{bmatrix}.
\]
From the first constraint we obtain \(\cos{\gamma/2}=\sin{\gamma/2} \Rightarrow \gamma=\pi/2\) and \(\beta=-\theta\).  
From the second constraint we find \(e^{i\delta}-e^{i\theta}=1\) and \(e^{i\delta}+e^{i\theta}=0\), which yield \(\delta=-\theta+\pi\).  

Thus, \(U\) can be decomposed into a sequence of \(R_x\) and \(R_y\) rotations. Using the \(\Pi(\phi,\theta,n)\) gate, these rotations can be made conditional on \(M \bmod 3\), ensuring that the Hadamard acts identically on the codespace and all correctable error spaces, thereby satisfying criterion~(ii) and preserving the noise bias.

\subsection{Phase Gates}

A logical phase gate \(P(\theta)\) on the 6-Cat code is implemented by mapping the logical states into the \(I_z\)-distinguishable basis, imprinting a controlled phase on the electron, and then mapping back to the original codespace. Choosing \(\theta = \pi/8\) implements the \(T\) gate.

The sequence begins with a CNOT (ensemble control) to entangle the logical state of the ensemble with the electron. The system then evolves freely in the rotated codespace for a duration \(t = \theta / \omega_e\), during which the electron’s \(\ket{\uparrow}\) component accumulates a phase \(e^{i\theta}\) relative to \(\ket{\downarrow}\). A conditional \(R_x(\pi, \Gamma)\) rotation on the electron disentangles it from the ensemble, transferring the acquired phase to the logical state. Finally, an ensemble rotation \(I_y(\pi/2)\) returns the state to the original codespace (Fig.~\ref{fig:gates_sq}d).

In the absence of errors, the action on the logical basis is:
\begin{align*}
\ket{0}_L &\rightarrow \ket{0}_L, \\
\ket{1}_L &\rightarrow e^{i\theta} \ket{1}_L,
\end{align*}
which is exactly the desired logical phase gate.  

If a raising or lowering error occurs before or during the gate, the ensemble acquires a nonzero magnetization, which would normally shift the electron precession frequency. This is avoided by making the conditional \(R_x(\pi,\Gamma)\) operation act on \(|0\rangle_L\) and all of its correctable error subspaces, ensuring that the acquired phase is applied consistently across the codespace and associated error spaces, in accordance with criterion~(ii). Because the ensemble polarization is restored at the end of the sequence, \(M \bmod \tfrac{N}{2}\) is preserved and the noise bias remains unchanged.

Thus, the \(P(\theta)\) gate imprints a controllable logical phase while maintaining full fault tolerance and bias preservation. The same construction applies for any \(\theta\), with \(\theta = \pi/8\) giving the logical \(T\) gate.

\bibliography{main.bbl}

\end{document}